\documentclass[a4paper,conference]{IEEEtran}
\IEEEoverridecommandlockouts
\usepackage{cite}
\usepackage{amsmath,amssymb,amsfonts}
\usepackage{algorithmic}
\usepackage[dvipdfmx]{graphicx}
\usepackage{subfigure}
\usepackage{textcomp}
\usepackage{xcolor}
\usepackage{url}
\def\BibTeX{{\rm B\kern-.05em{\sc i\kern-.025em b}\kern-.08em
    T\kern-.1667em\lower.7ex\hbox{E}\kern-.125emX}}
\begin{document}
\title{Single-Shot High Dynamic Range Imaging with Spatially Varying Exposures Considering Hue Distortion}

\author{\IEEEauthorblockN{Chihiro Go, Yuma Kinoshita, Sayaka Shiota and Hitoshi Kiya}
\IEEEauthorblockA{\textit{Tokyo Metropolitan University}, Tokyo, Japan\\
E-mail: go-chihiro@ed.tmu.ac.jp, kinoshita-yuma@ed.tmu.ac.jp, sayaka@tmu.ac.jp, kiya@tmu.ac.jp}
}

\maketitle

\begin{abstract}
We proposes a novel single-shot high dynamic range imaging scheme with spatially varying exposures (SVE) considering hue distortion.
Single-shot imaging with SVE enables us to capture multi-exposure images from a single-shot image, so high dynamic range images can be produced without ghost artifacts. 
However, SVE images have some pixels at which a range supported by camera sensors is exceeded.
Therefore, generated images have some color distortion, so that conventional imaging with SVE has never considered the influence of this range limitation.
To overcome this issue, we consider estimating the correct hue of a scene from raw images, and propose a method with the estimated hue information for correcting the hue of SVE images on the constant hue plain in the RGB color space.
\end{abstract}


\section{Introduction}
The low dynamic range (LDR) imaging sensors used in modern digital cameras cannot express the dynamic range of a real scene, due to a limited dynamic range which imaging sensors have \cite{kinoshita2018automatic,kinoshita2018pseudo,kinoshita2019scene}.
The limitation results in the low contrast of images taken by digital cameras.
The most common approach for HDR imaging is to fuse multi-exposure images which are to merge a set of LDR images taken with different exposure times.
This approach requires to capture multi-exposure images by taking at the different time, so there are ghost artifact issues, due to the movement of the camera and the subject.
One of ghost-free techniques for HDR imaging is to employ spatially varying exposures (SVE)\cite{alex,cho2014single,hajsharif2014hdr,gil2016high}.
In  the SVE-based imaging, a scene is captured with varying exposures for each pixel in a single image, and multiple sub-images with each exposure are obtained.
However, conventional SVE-based methods focus on the luminance of a scene, so they cause color distortion, due to the influence of the limited dynamic range.

To overcome this issue, a novel single-shot imaging scheme with SVE is proposed in this paper.
The correct hue of a scene is estimated from raw images, and then the estimated hue information is employed on the constant hue plain in the RGB color space \cite{ueda2018hue} for correcting the hue of SVE images.
\section{Related works}
\subsection{SVE image}
\begin{figure}[tp]
	\centering
	\includegraphics[width = 0.90\columnwidth]{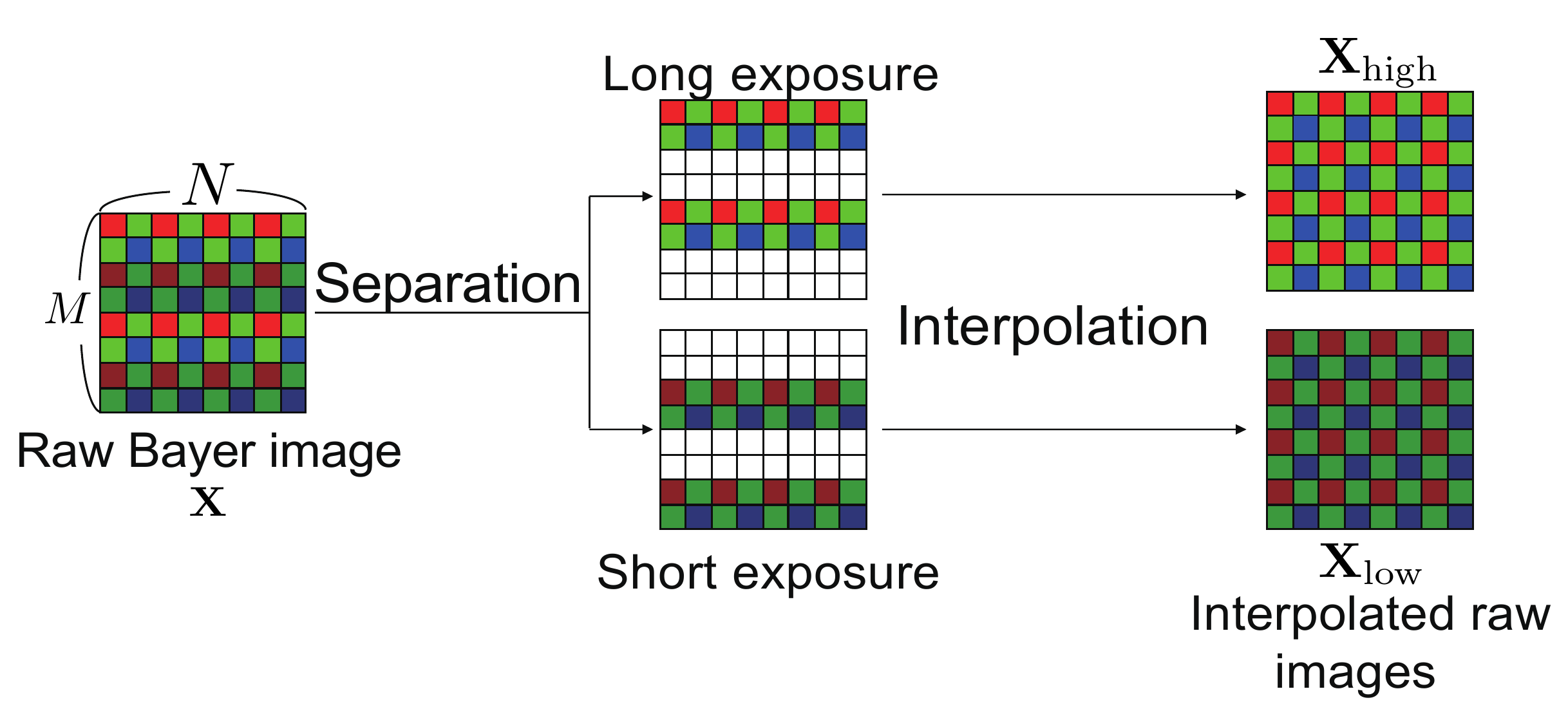}
	\caption{SVE image}
	\label{fig:sve_sensor}
\end{figure}

A raw Bayer image ${\bf X}$ sensed with SVE sensor is illustrated in Fig.\ref{fig:sve_sensor}, where the exposure value alternates every two lines in the Bayer image.
The raw image $\bf X$ is separated into two images according to exposure values.
An interpolation operation is applied to each raw image for producing two raw images with the same size as $\bf X$: ${\bf X}_{\rm low}$ and ${\bf X}_{\rm high}$.
However, $\bf X$ has some pixels at which a range supported by the camera sensor is exceeded.
\subsection{Constant hue place in the RGB color space}
\begin{figure}[tp]
	\centering
	\includegraphics[width = 0.90\columnwidth]{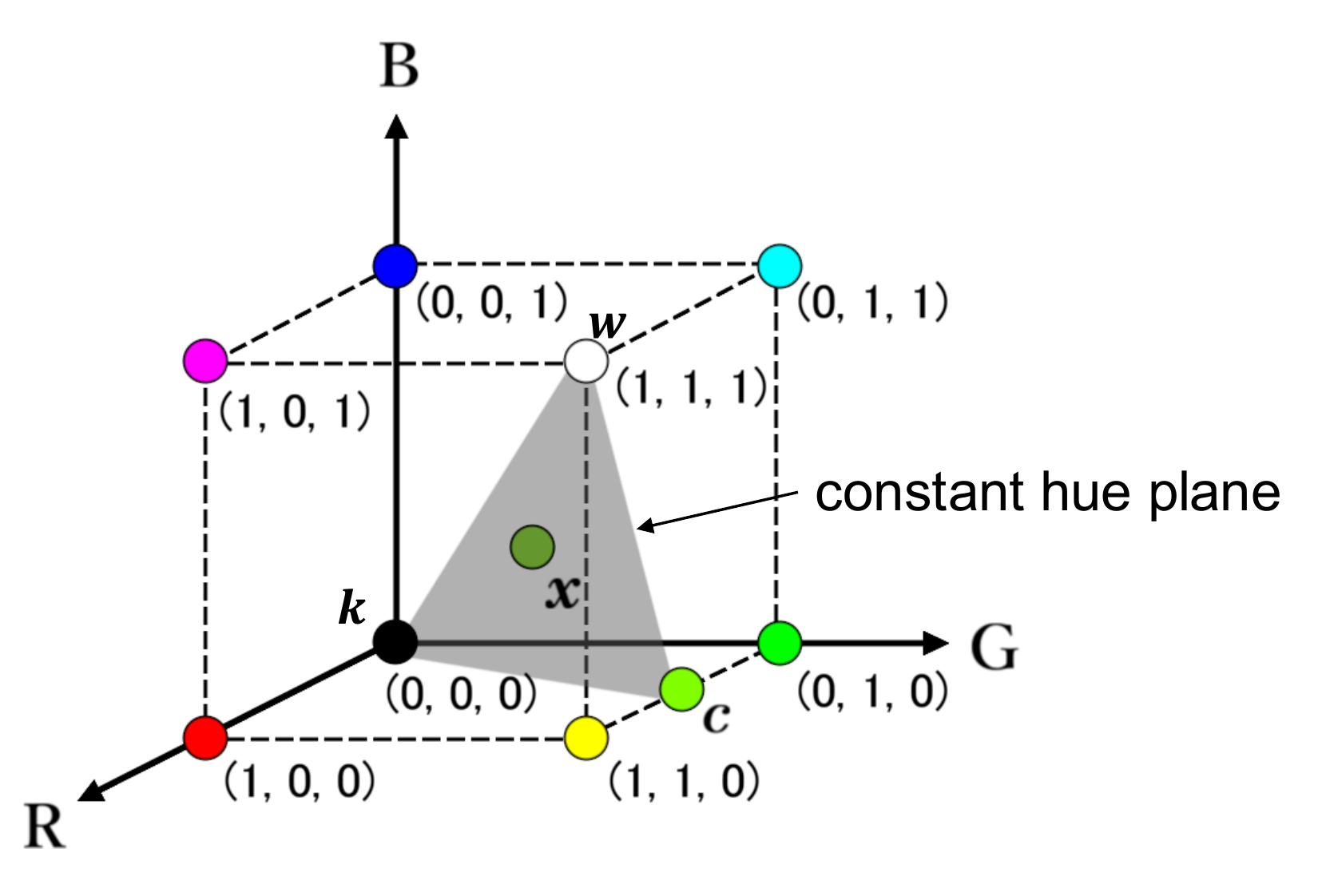}
	\caption{Constatnt hue plane with RGB color space}
	\label{fig:color_space}
\end{figure}

An input image is a 24-bit full color image and each pixel of the image is represented as $\mbox{\boldmath $x$} \in [0,1]^3$.
$x_r, x_g$ and $x_b$ are the R, G, and B components of the pixel $\mbox{\boldmath $x$}$, respectively, as shown in Fig.\ref{fig:color_space}.
In the RGB color space, a set of pixels which has the same hue forms a plane, called constant hue plane \cite{ueda2018hue}.
The shape of the constant hue plane is the triangle whose vertices correspond to white, black and the maximally saturated color, where $\mbox{\boldmath $w$} = (1,1,1)$, $\mbox{\boldmath $k$} = (0,0,0)$ and $\mbox{\boldmath $c$}$ are white, black and the maximally saturated color with the same hue as $\mbox{\boldmath $x$}$, respectively.
The maximally saturated color $\mbox{\boldmath $c$} = (c_r,c_g,c_b)$ is calculated by, under $l = \{r,g,b\}$
\begin{equation}
	c_l = \frac{x_l-\min(\mbox{\boldmath $x$})}{\max(\mbox{\boldmath $x$})-\min(\mbox{\boldmath $x$})}
	\label{eq:c_r,g,b}
\end{equation}
where $\max(\cdot)$ and $\min(\cdot)$ are functions that return the maximum and minimum elements of the pixel \mbox{\boldmath $x$}, respectively.

A pixel $\mbox{\boldmath $x$}$ can also be represented as a linear combination as
\begin{equation}
	\mbox{\boldmath $x$} = a_w\mbox{\boldmath $w$}+a_k\mbox{\boldmath $k$}+a_c\mbox{\boldmath $c$}
	\label{eq:x_wkc}
\end{equation}
where the coefficient meet the equations,
\begin{equation}
	a_w+a_k+a_c=1,
\end{equation}
\begin{equation}
	0\leq a_w,a_k,a_c\leq1.
\end{equation}
This method is applied to various methods \cite{kobayashi2019jpeg,visavakitcharoen2019pure}.

\section{Proposed method}
\begin{figure}[tp]
	\centering
	\includegraphics[width = 0.90\columnwidth]{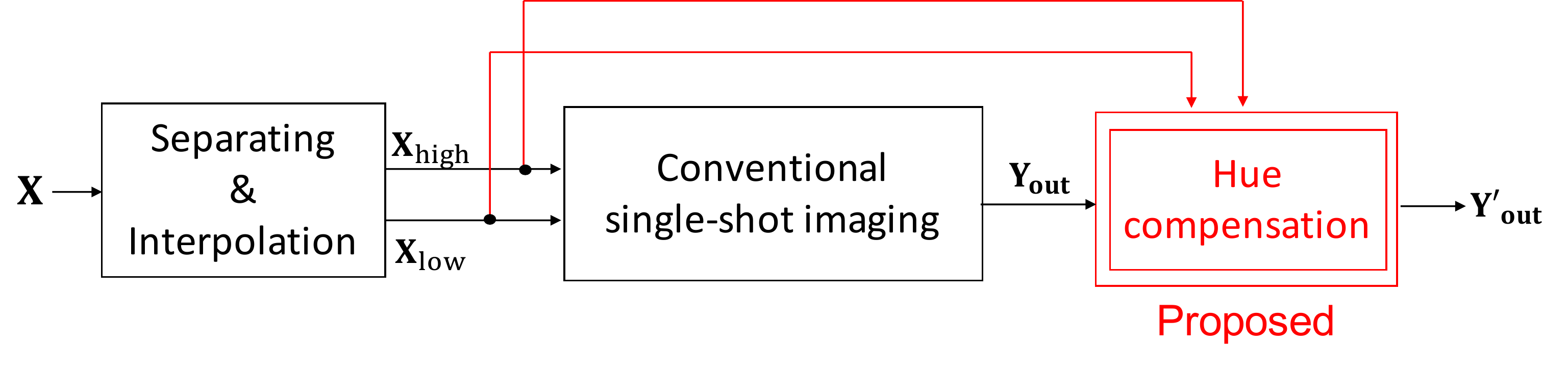}
	\caption{Outline of proposed method}
	\label{fig:proposed_algorithm}
\end{figure}
The outline of the proposed method is shown in Fig.\ref{fig:proposed_algorithm}.
\subsection{Procedure}
\subsubsection{Separation and interpolation}
A raw image $\bf X$ is first divided into two raw images, according to the exposure value.
Next, interpolation processing is applied to each raw image for producing two raw images: ${\bf X}_{\rm low}$ and ${\bf X}_{\rm high}$

\subsubsection{Exposure compensation}
Perform scene-segmentation based exposure compensation \cite{kinoshita2019scene}.

\subsubsection{Demosaicing}
A demosaicing algorithm is applied to the compensated images to obtain RGB images. 

\subsubsection{Image fusion}
The RGB images are fused by using a fuse function.

\subsubsection{Hue compensation}
The hue of the fused image ${\bf Y}_{\rm out}$ is compensated by using the proposed method as shown below.
\subsection{Hue estimation and hue correction}
From Eq.\eqref{eq:x_wkc}, a pixel value $\mbox{\boldmath $y$}_{\rm out}$ in ${\bf Y}_{\rm out}$ is given by,
\begin{equation}
	\mbox{\boldmath $y$}_{out} = a_w\mbox{\boldmath $w$}+a_k\mbox{\boldmath $k$}+a_c\mbox{\boldmath $c$}_{\rm out},
\end{equation}
where $\mbox{\boldmath $c$}_{\rm out}$ is the maximally saturated color of $\mbox{\boldmath $y$}_{out}$, and $a_w$, $a_k$, $a_c$ and $\mbox{\boldmath $c$}_{out}$ are calculated from $\mbox{\boldmath $y$}_{out}$.
Note that $\mbox{\boldmath $c$}_{out}$ may be  distorted due to the influence of some pixels at which a range supported by the camera sensor is exceeded.
Each pixel value $\mbox{\boldmath $y$}$ of the conventional method is recalculated using the maximally saturated color values $\mbox{\boldmath $c$}_{\rm low},\mbox{\boldmath $c$}_{\rm high}$ calculated form each pixel value $\mbox{\boldmath $x$}_{\rm low},\mbox{\boldmath $x$}_{\rm high}$ to suppress the hue distorition.
Therefore, we propose replacing $\mbox{\boldmath $c$}_{out}$ with $\mbox{\boldmath $c$}'_{out}$ to reduce the hue distortion, as,
\begin{equation}
	\mbox{\boldmath $y$}'_{out} = a_w\mbox{\boldmath $w$}+a_k\mbox{\boldmath $k$}+a_c\mbox{\boldmath $c$}'_{\rm out},
\end{equation}
where
\footnotesize
\begin{equation}
	\mbox{\boldmath $c$}'_{\rm out} =
	\begin{cases}
		\mbox{\boldmath $c$}_{\rm low} & {\rm if}\; x_{\rm low}\neq 0\;{\rm and}\;x_{\rm low}\neq1\\
		\mbox{\boldmath $c$}_{\rm high} & {\rm if}\;x_{\rm high}\neq 0,\;x_{\rm high}\neq1,\;x_{\rm low}= 0,\;{\rm and}\;x_{\rm low}=1\\
		\mbox{\boldmath $c$}_{\rm out} & {\rm if}\; x_{\rm low}= 0,\;x_{\rm low}=1,\;x_{\rm high}= 0,\;{\rm and}\;x_{\rm high}=1\\
	\end{cases}
	\label{eq:proposed}
\end{equation}
\normalsize
where $x_{\rm low}$ and $x_{\rm high}\in[0,1]$ are pixel values in the raw images, ${\bf X}_{\rm low}$ and ${\bf X}_{\rm high}$, and the maximum saturated colors $\mbox{\boldmath $c$}_{\rm low}$ and $\mbox{\boldmath $c$}_{\rm high}\in[0,1]^3$ are calculated from $x_{\rm low},x_{\rm high}$.
RGB images, ${\bf Y}_{\rm low}$ and ${\bf Y}_{\rm high}$ are first calculated by applying a demosaicing algorithm to ${\bf X}_{\rm low}$ and ${\bf X}_{\rm high}$, and then $\mbox{\boldmath $c$}_{\rm low}$ and $\mbox{\boldmath $c$}_{\rm high}$ are calculated from pixel values $\mbox{\boldmath $y$}_{\rm low}$ and $\mbox{\boldmath $y$}_{\rm high}$ of ${\bf Y}_{\rm low}$ and ${\bf Y}_{\rm high}$, respectively.
Eq.\eqref{eq:proposed} allows us to is to avoid using the pixels at which over-exposure and under-exposure occurs.

\section{Experiment}
In an experiment, the performance of the proposed scheme was compared with the conventional single-shot imaging with SVE .
\subsection{Dataset}
564 input SVE images $\bf X$ were prepared by using 141 HDR images selected from a database \cite{hdrlabs}.
Four SVE image sets with two exposure values $\pm 1$EV, $\pm 2$EV, $\pm3$EV, or $\pm4$EV were generated as ${\bf X}$ from each HDR image.

\subsection{Objective metrics}
The hue distortion of images produced by each method was evaluated in two objective metrics; the cosine similarity of maximally saturated color values, and the difference of hue values in CIEDE2000 \cite{luo2001development}.
The difference of hue values between a reference image (HDR) and the generated one was calculated for each pixel, and then the average value of all pixels was computed.
For cosine similarity, a larger value means higher quality, and for the difference of hue values, a smaller value means higher quality.

The quality of images produced by each method was evaluated in the objective metrics; the tone mapped image quality index (TMQI) \cite{yeganeh2013objective}.
TMQI measure the quality of a tone mapped image from an HDR image and it consists of structural fidelity and statistical naturalness.
For TMQI, a larger value means higher quality.

\subsection{Experiment results}
From Table \ref{tb:pc_all} and Table \ref{tb:ciede_all}, it is confirmed that the proposed method had higher scores than conventional method. 
Therefore, the proposed method is effective for improving hue distortion.

From Table \ref{tb:tmqi_all}, it is confirmed that the proposed method had lower scores than conventional method. 
Although the image quality decreases slightly, the performance of the conventional method can be maintained in terms of TMQI.
\begin{table}[tp]
  \center
  \caption{Average scores of the maximally saturated color similarity}
  \begin{tabular}{| l | c | r | r | r |} \hline
    					& $\pm 1{\rm EV}$ & $\pm 2{\rm EV}$ & $\pm 3{\rm EV}$ & $\pm 4{\rm EV}$ \\ \hline \hline
    Conventional method	&0.9250&0.9290&0.9308&0.9304\\
    Proposed method		&\bf0.9302&\bf0.9382&\bf0.9432&\bf0.9441\\ \hline
  \end{tabular}
  \label{tb:pc_all}
\end{table}
\begin{table}[tp]
  \center
  \caption{Average scores of the difference hue values in CIEDE2000}
  \begin{tabular}{| l | c | r | r | r |} \hline
    					& $\pm 1{\rm EV}$ & $\pm 2{\rm EV}$ & $\pm 3{\rm EV}$ & $\pm 4{\rm EV}$ \\ \hline \hline
    Conventional method	&15.15&14.73&15.07&15.28\\
    Proposed method		&\bf15.00&\bf14.53&\bf14.87&\bf15.23\\ \hline
  \end{tabular}
  \label{tb:ciede_all}
\end{table}
\begin{table}[tp!]
  \center
  \caption{Average scores of TMQI}
  \begin{tabular}{| l | c | r | r | r |} \hline
    					& $\pm 1{\rm EV}$ & $\pm 2{\rm EV}$ & $\pm 3{\rm EV}$ & $\pm 4{\rm EV}$ \\ \hline \hline
    Conventional method	&\bf0.2126&\bf0.2120&\bf0.2106&\bf0.2091\\
    Proposed method		&0.2121&0.2115&0.2101&0.2084\\ \hline
  \end{tabular}
  \label{tb:tmqi_all}
\end{table}

\section{Conclusion}
In this paper, we proposed a novel single-shot high dynamic range imaging scheme with SVE considering hue distortion.
We considered estimating the correct hue of a scene from raw images, and proposed a method with the estimated hue information for correcting the hue of SVE images on the constant hue plain in the RGB color space.

\bibliographystyle{IEEEtran}

\begin{thebibliography}{00}
\bibitem{kinoshita2018automatic}
Y.~Kinoshita and H.~Kiya, ``Automatic exposure compensation using an image segmentation method for single-image-based multi-exposure fusion,'' \emph{APSIPA Transactions on Signal and Information Processing}, vol.~7, 2018.

\bibitem{kinoshita2018pseudo}
Y.~Kinoshita, S.~Shiota, and H.~Kiya, ``A pseudo multi-exposure fusion method using single image,'' \emph{IEICE Transactions on Fundamentals of Electronics, Communications and Computer Sciences}, vol. 101, no.~11, pp. 1806--1814, 2018.

\bibitem{kinoshita2019scene}
Y.~Kinoshita and H.~Kiya, ``{Scene Segmentation-Based Luminance Adjustment for Multi-Exposure Image Fusion},'' \emph{IEEE Trans. Image Processing.}, vol.~28, no.~8, pp. 4101--4116, August 2019.

\bibitem{alex}
A1EX, ``Dynamic range improvement for some canon dslrs by alternating iso during sensor readout,'' \url{http://acoutts.com/a1ex/dual_iso.pdf}, 2013.

\bibitem{cho2014single}
H.~Cho and S.~Lee, ``Single-shot high dynamic range imaging using coded electronic shutter,'' in \emph{Computer Graphics Forum}, vol.~33, July 2013.

\bibitem{hajsharif2014hdr}
S.~Hajsharif, J.~Kronander, and J.~Unger, ``{HDR} reconstruction for alternating gain ({ISO}) sensor readout,'' April 2014.

\bibitem{gil2016high}
R.~Gil~Rodr{\'\i}guez and M.~Bertalm{\'\i}o, ``High quality video in high dynamic range scenes from interlaced dual-iso footage,'' The Society for Imaging Science and Technology (IS\&T), 2016.

\bibitem{ueda2018hue}
U.~Yoshiaki, M.~Hideaki, K.~Takanori, and E.~U. Noriaki~Suetake, ``Hue-preserving color contrast enhancement method without gamut problem by using histogram specification,'' in \emph{Proceeding ICIP}.\hskip 1em plus 0.5em minus 0.4em\relax IEEE, 2018, pp. 1123--1127.

\bibitem{kobayashi2019jpeg}
H.~Kobayashi and H.~Kiya, ``Jpeg xt image compression with hue compensation for two-layer hdr coding,'' \emph{arXiv preprint arXiv:1904.11315}, 2019.

\bibitem{visavakitcharoen2019pure}
A.~Visavakitcharoen, Y.~Kinoshita, and H.~Kiya, ``Pure-color preserving multi-exposure image fusion,'' in \emph{International Workshop on Advanced Image Technology (IWAIT) 2019}, vol. 11049.\hskip 1em plus 0.5em minus 0.4em\relax International Society for Optics and Photonics, 2019, p. 110493X.

\bibitem{hdrlabs}
hdrlabs, ``s{IBL} {A}rchive,'' http://www.hdrlabs.com/sibl/archive.html.

\bibitem{luo2001development}
M.~R. Luo, G.~Cui, and B.~Rigg, ``The development of the cie 2000 colour-difference formula: Ciede2000,'' \emph{Color Research \& Application}, vol.~26, no.~5, pp. 340--350, 2001.

\bibitem{yeganeh2013objective}
H.~Yeganeh and Z.~Wang, ``Objective quality assessment of tone-mapped images,'' \emph{IEEE Transactions on Image Processing}, vol.~22, no.~2, pp. 657--667, 2013.
  \end{thebibliography}

\end{document}